\documentclass[USenglish]{eptcs}

\usepackage{amssymb}
\usepackage{amsmath}
\usepackage{amsthm}
\usepackage{bussproofs}
\usepackage{caption}
\usepackage[noadjust]{cite}
\usepackage{eufrak}
\usepackage{graphicx}
\usepackage{mdwlist}
\usepackage{subfig}
\usepackage{ulem}
\usepackage{listings}
\usepackage{url}
\usepackage{tikz}
\usetikzlibrary{positioning,fit,arrows,decorations.markings}
\usepackage{xspace}

\definecolor{funsym}{rgb}{0.0,0.0,0.75}
\definecolor{varsym}{rgb}{0.75,0.0,0.0}
\newsavebox{\cpfbox}
\newenvironment{cpfsnip}%
  {\begin{lrbox}{\cpfbox}\begin{minipage}{0.95\textwidth}}%
  {\end{minipage}\end{lrbox}\begin{center}\fbox{\usebox{\cpfbox}}\end{center}}

\newcommand\aprove{\textsf{AProVE}\xspace}
\newcommand\matchbox{\textsf{Matchbox}\xspace}
\newcommand\cime{\textsf{C\kern-0.1ex\textsl{i}ME}\xspace}
\newcommand\tttt{%
 \textsf{T\kern-0.2em\raisebox{-0.3em}T\kern-0.2emT\kern-0.2em\raisebox{-0.3em}2}\xspace}
\newcommand{\csi}{\textsf{CSI}\xspace}
\newcommand\tct{%
 \textsf{T\kern-0.2em\raisebox{-0.3em}C\kern-0.2emT}\xspace}
\newcommand{\cat}{\textsf{C\raisebox{0.15em}a\kern-0.15emT}\xspace}
\newcommand{\kbcv}{\textsf{KBCV}\xspace}
\newcommand{\mkbtt}{\textsf{mkbTT}\xspace}

\DeclareSymbolFont{letters}{OML}{cmbboard}{m}{it} 

\newcommand{\from}{\leftarrow}
\def\test#1#2#3{\setbox0=\hbox{$\vphantom{#1}^{#2}_{#3}$}%
                \dimen0=\wd0%
                \setbox1=\hbox{$\scriptstyle #2$}%
                \advance\dimen0-\wd1%
                \setbox1=\hbox{\hskip\dimen0\copy1}%
                \dimen0=\wd0%
                \setbox2=\hbox{$\scriptstyle #3$}%
                \advance\dimen0-\wd2%
                \setbox2=\hbox{\hskip\dimen0\copy2}%
                {\vphantom{#1}^{\box1}_{\box2}}{#1}
}

\newcommand\EE{\ensuremath{\mathcal{E}}\xspace}

\newcommand\RR{\ensuremath{\mathcal{R}}\xspace}

\newcommand{\rSC}[1]{Section~\ref{#1}}

\newcommand{\rFI}[1]{Figure~\ref{#1}}

\newcommand{\sep}{\hspace{-0.28em}}

\newlength{\rtimesl}
\newcommand{\overlay}%
{\mathrel{\from\sep{\ltimes\hspace{-\rtimesl}\rtimes}\sep\to}}

\newcommand\rainbow{\textsf{Rainbow}\xspace}
\newcommand\colorr{\textsf{CoLoR}\xspace}
\newcommand\CoLoR\colorr
\newcommand\cocci{\textsf{Coccinelle}\xspace}

\newcommand\isafor{\textsf{Isa\kern-0.2exF\kern-0.2exo\kern-0.2exR}\xspace}
\newcommand\ceta{\textsf{C\kern-0.2exe\kern-0.5exT\kern-0.5exA}\xspace}
\newcommand\colrain{\colorr/\rainbow}
\newcommand\coccime{\cocci/\cime}
\newcommand\isaceta{\isafor/\ceta}

\title{The Certification Problem Format%
\footnote{%
  This research is supported by the Austrian Science Fund (FWF) projects P22767
  and J3202.}}

\def\titlerunning{The Certification Problem Format}
\def\authorrunning{C.\ Sternagel and R.\ Thiemann}

\author{Christian Sternagel 
\institute{University of Innsbruck, Austria}
\email{christian.sternagel@uibk.ac.at}\and 
Ren{\'e} Thiemann
\institute{University of Innsbruck, Austria}
\email{rene.thiemann@uibk.ac.at}}

\newcommand\groupop{\bullet}
\newcommand\groupid{\mathsf{e}}
\newcommand\groupinv[1]{#1^-}
\newcommand\hbull{\textcolor{funsym}{\text{$\bullet$}}}
\newcommand\hvar[1]{\textcolor{varsym}{\text{#1}}}
\newcommand\hid{\textcolor{funsym}{\text{e}}}
\newcommand\hinv{\textcolor{funsym}{\text{-}}}

\newcommand\hi[1]{\par\textbf{\Large#1}\par}
\newcommand\hii[1]{\par\smallskip\textbf{\large#1}\par}
\newcommand\hiii[1]{\par\smallskip\textbf{#1}\par}

\newcommand\figref[1]{Figure~\ref{fig:#1}}

\newtheorem{example}{Example}

\begin{document}
\maketitle

\begin{abstract}
We provide an overview of CPF, the certification problem format, and explain some design decisions.
Whereas CPF was originally invented
to combine three different formats for termination proofs into a single 
one, in the meanwhile proofs for several other properties of term rewrite systems are
also expressible: like confluence, complexity, and completion. As a consequence, 
the format is already supported by several tools and certifiers. Its acceptance is also demonstrated
in international competitions: the certified tracks of both the termination and the 
confluence competition utilized CPF as exchange format between automated tools and 
trusted certifiers. 
\end{abstract}

\section{Introduction}

Automated tools that perform logical deductions are available in several areas.
For example,
there are SAT solvers, SMT solvers, automated theorem provers for first-order
logic (FTP), termination tools, complexity tools, confluence tools, completion tools, etc.
In most areas, the community was able to agree on a \emph{common} input format,
like the DIMACS, TPTP, and TPDB formats. Such a format is beneficial for
several reasons: for example, users can easily try different
tools on their problems and it is possible to compare tools by 
running experiments on large databases of problems. 

One problem when using such automated tools is that they are
complex pieces of software, and thus may contain bugs. These bugs may be
harmless---%
e.g., the tool just crashes or does not provide an answer where in principle
it should be able to find one---or in the worst case
lead to wrong answers. For this reason, certification of the generated answers
is an important task. 

Of course, to certify an answer, the result of an automated deduction tool must
not be just a simple yes/no. Instead it has to provide a sufficiently
detailed proof for validating the answer.  Most of these proofs
can be seen as a composition of several basic proof steps. But there are
exceptions: for SAT and SMT, a proof of satisfiability can be given by just 
providing the satisfying assignment, so here no compositional proof is required.

In the following we shortly discuss some differences of the structure
of these compositional proofs. 

\begin{itemize}
\item \emph{Complexity of basic proof steps}:
  A proof of unsatisfiability for SAT can be performed in various 
  frameworks (natural deduction, resolution, DPLL), 
  which all have very simple inference rules. 
  Also for FTP,
  the basic proof steps are rather easy (natural deduction, 
  resolution, superposition, basic step in completion procedure). 
  In contrast, basic proof steps in SMT solvers can be complex (apply decision procedures
  for supported theories) and also for termination, confluence or complexity proofs of 
  term rewrite systems (TRSs) a single
  proof step can be complex.
  For example, in the termination technique called semantic labeling \cite{Z95}
  one has to check that the given interpretation forms a model; for determining 
  complexity of some TRS via matrix interpretations \cite{EWZ08} one has to estimate
  growth rates of matrix-products, 
  and for disproving confluence one has to ensure non-reachability under rewriting. 

\item \emph{Number of basic proof steps in a compositional proof}:
  Complexity, confluence, and termination proofs often require only a small number
  of proof steps in comparison to the number of steps within proofs for SAT, SMT, and FTP.

\item \emph{Set of inference rules}:
  The set of inference rules that are used for tools in the 
  areas SAT, SMT, and FTP is rather static---the inference rules are 
  fixed by the
  respective frameworks of natural deduction, resolution, etc.
  In contrast, the set of techniques that are used in confluence, complexity, and termination tools 
  is often dynamic---much of the
  power of these tools relies on the invention of new ways to
  prove these properties, e.g., by inventing new kinds of well-founded orders, etc.
  
\item \emph{Determinism of basic proof steps}:
  Several proof steps are completely determined, like a conjunction introduction within 
  natural deduction. But there are also basic proof steps that need further
  information to determine the result. For example, from one conflict in DPLL one
  can learn different conflict clauses.
\end{itemize}

To summarize, proofs for TRSs (with properties like termination, confluence, and
complexity) are usually small in terms of number of steps, but basic proof steps
may be complex. Moreover, the set of available basic proof steps is constantly
growing. 

In this paper, we present the certification problem
format (CPF), a format initially developed to represent termination proofs for TRSs which
has recently been extended to also support confluence, complexity, and completion proofs.
It has four major benefits.
First, it is easy for automated tools to generate CPF files; second,
it is easy to add new techniques to CPF; third, it provides enough
information for certification; finally, it is 
a \emph{common} proof format that is supported by several tools and 
certifiers.
The last point is also why the word \emph{problem} is part of the name CPF: In
the certified categories of the termination competition, first several tools
produce CPF certificates, which are then used as input \emph{problems} for
different certifiers.

All details on CPF and several example proofs are freely available at the following URL:
\begin{center}
\url{http://cl-informatik.uibk.ac.at/software/cpf/}
\end{center}
The main file is \texttt{cpf.xsd}, the detailed schema for CPF,
which can be seen as an algebraic
datatype for proofs and which formally specifies all the information that has to be provided
in each proof step, in combination with clarifying comments.

\section{An Example from Group Theory}
\label{example}
Before going into a more detailed explanation of CPF, let us consider a concrete
example. As already mentioned above, two of the main properties that are covered by the
format are termination and confluence (of term rewrite systems). A well-known
technique whose certification requires a combination of both of these properties
is (Knuth-Bendix) completion~\cite{KB70}. That is, given a set of equations, the goal is to
obtain a terminating and confluent rewrite system that serves as a decision
procedure for the word problem (i.e., the question whether two terms are equal
with respect to the given equations). The prime example is group theory. More
precisely, given the three group axioms
\begin{align*}
(x \groupop y) \groupop z &= x \groupop (y \groupop z)
  & &\text{(associativity)}\\
\groupid \groupop x &= x
  & &\text{(left neutral)}\\
\groupinv{x} \groupop x &= \groupid
  & &\text{(left inverse)}
\end{align*}
find a set of rewrite rules, such that two given terms are equal with respect to
the group axioms if and only if an exhaustive application of these rules to both
of them, leads to syntactic equality. By termination of the rules an exhaustive
application is always possible (since we will hit a normal form, i.e., a term to
which none of the rules are applicable anymore, eventually), while confluence
guarantees that we reach the same normal form independent of the employed
rewriting strategy.

In the following we present a CPF certificate (as printed in a web browser, in
order to increase readability) for the group theory example. As shown in
\figref{ini}, the certificate starts by stating the kind of proof under
consideration as well as the specific input (in this case a \emph{completion
proof} for the group equations). Furthermore, also the resulting TRS is part of
the input. (Both are given in prefix notation, i.e., $\hbull(\hvar{x},
\hvar{y})$ instead of $x \groupop y$ and $\hinv(\hvar{x})$ instead of
$\groupinv{x}$.)

\begin{figure}
\begin{cpfsnip}
\hi{Completion Proof}
by kbcv (version 1.7)
\hii{Input}
For the following equations E
\begin{align*}
\hbull(\hbull(\hvar{x}, \hvar{y}), \hvar{z}) &=
  \hbull(\hvar{x},\hbull(\hvar{y},\hvar{z})) \\
\hbull(\hid, \hvar{x}) &= \hvar{x}\\
\hbull(\hinv(\hvar{x}), \hvar{x}) &= \hid
\end{align*}
and the following TRS R
\begin{align*}
\hbull(\hbull(\hvar{x},\hvar{y}),\hvar{z}) &\to \hbull(\hvar{x},\hbull(\hvar{y},\hvar{z}))\\
\hbull(\hid,\hvar{x})  &\to \hvar{x}\\
\hbull(\hinv(\hvar{x}),\hvar{x}) &\to \hid\\
\hbull(\hinv(\hvar{x}),\hbull(\hvar{x},\hvar{z}))  &\to \hbull(\hid,\hvar{z})\\
\hinv(\hid)  &\to \hid\\
\hbull(\hvar{x},\hid)  &\to \hvar{x}\\
\hinv(\hinv(\hvar{x})) &\to \hvar{x}\\
\hbull(\hvar{x},\hinv(\hvar{x})) &\to \hid\\
\hbull(\hvar{x},\hbull(\hinv(\hvar{x}),\hvar{z}))  &\to \hvar{z}\\
\hinv(\hbull(\hvar{y},\hvar{x})) &\to \hbull(\hinv(\hvar{x}),\hinv(\hvar{y}))
\end{align*}
it is proven that E is equivalent to R, and R is convergent.
\end{cpfsnip}
\caption{\label{fig:ini}Input specification of group theory certificate}
\end{figure}

\begin{figure}
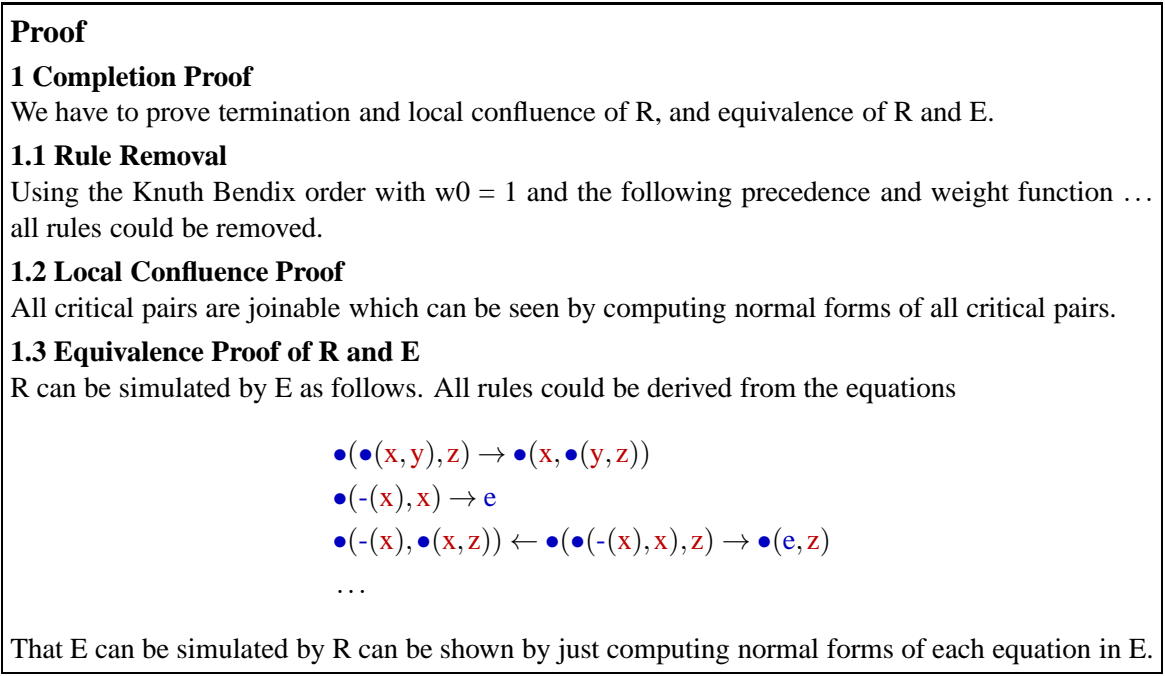

\begin{cpfsnip}
\hii{Proof}
\hiii{1 Completion Proof}
We have to prove termination and local confluence of R, and equivalence of R and
E.
\hiii{1.1 Rule Removal}
Using the Knuth Bendix order with w0 = 1 and the following precedence and weight
function $\ldots$ all rules could be removed.
\hiii{1.2 Local Confluence Proof}
All critical pairs are joinable which can be seen by computing normal forms of
all critical pairs.
\hiii{1.3 Equivalence Proof of R and E}
R can be simulated by E as follows. All rules could be derived from the
equations
\begin{align*}
& \hbull(\hbull(\hvar{x},\hvar{y}),\hvar{z}) \to \hbull(\hvar{x},\hbull(\hvar{y},\hvar{z})) \\
& \hbull(\hinv(\hvar{x}),\hvar{x}) \to \hid \\
& \hbull(\hinv(\hvar{x}),\hbull(\hvar{x},\hvar{z})) \from \hbull(\hbull(\hinv(\hvar{x}),\hvar{x}),\hvar{z}) \to \hbull(\hid,\hvar{z}) \\
& \ldots
\end{align*}
That E can be simulated by R can be shown by just computing normal forms of
each equation in E.
\end{cpfsnip}
\caption{\label{fig:proof}Proof part of group theory certificate}
\end{figure}

The remainder of the certificate (\figref{proof}) contains the corresponding
proof, or at least enough information that such a proof can be reconstructed by
a certifier. For brevity, some parts of the proof are omitted (indicated by
$\ldots$).

As stated in the certificate, for proving that the given rewrite system (R)
is indeed a correct completion result, there are two obligations.
First, we have to show termination and local
confluence of R (the combination of which also yields confluence, by Newman's lemma).
In addition, we have to make sure that R is equivalent to E in the sense that
two terms are equal with respect to the equations in E if and only if one can be
transformed into the other by applying rules of R in an arbitrary order and in
arbitrary directions.

\section{The Certification Problem Format}

Prior to the development of certifiers for termination proofs, each termination
tool provided some kind of human readable justification for its result, e.g., a
plain text or HTML description of the applied techniques.  It is hard to extract
the relevant proof steps from this kind of justification, since parameters of
termination techniques are mixed with human readable explanations. Moreover, the
output was not standardized at all, i.e., every tool had its own variant. 

In this setting it is not surprising that the first certifiers for
termination proofs (\colrain~\cite{color}, \coccime~\cite{a3pat-cime}, and later
\isaceta~\cite{ceta-tphols}) each expected their own input format.  Hence, to
support certifiable proofs for all certifiers, a termination tool had to provide
several proof output routines, as illustrated in \rFI{before}. 

\tikzstyle{tool}=[transform shape,shape = rectangle, draw=black,fill=white,text width=1.8cm,text centered] 
\tikzstyle{toolc}=[transform shape,shape = rectangle, draw=black,fill=white,text width=2.8cm,text centered] 
\tikzstyle{cert}=[transform shape,shape = rectangle, draw=black,fill=white,text width=2.8cm,text centered] 

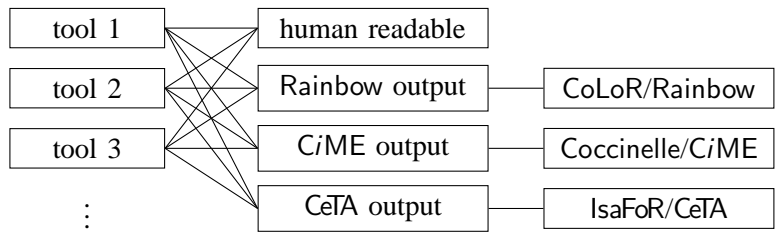
\begin{figure}
\centering
\begin{tikzpicture}[node distance = 0.8cm]
\node[tool] (1) {tool 1};
\node[tool, below of = 1] (2) {tool 2};
\node[tool, below of = 2] (3) {tool 3};
\node[below of = 3] (n) {$\vdots$};
\node[cert, right of = 1,xshift=3cm] (a) {human readable};
\node[cert, below of = a] (b) {\rainbow output};
\node[cert, below of = b] (c) {\cime output};
\node[cert, below of = c] (d) {\ceta output};
\node[toolc, right of = b, xshift = 3cm] (B) {\colrain};
\node[toolc, below of = B] (C) {\coccime};
\node[toolc, below of = C] (D) {\isaceta};
\foreach \x in {1,2,3}
   {\foreach \y in {a,b,c,d} {\draw (\x.east) -- (\y.west);}}
\draw (b) -- (B);
\draw (c) -- (C);
\draw (d) -- (D);
\end{tikzpicture}
\caption{\label{before}Certification of proofs before CPF}
\end{figure}

In order to reduce the burden for termination tool authors, the development
teams of the three certifiers decided to establish \emph{a single} proof
format that should be supported by all certifiers, namely CPF.  As a result,
nowadays termination tools only have to generate CPF certificates independent of
the intended certifier.  Of course, also the feedback of termination tool
authors was considered during the development of CPF. By now it is widely
accepted in the community and in fact the only format that is used in the
certified categories of the termination competition. 

Originally CPF was designed solely for termination proofs. Nowadays, it can also
be used for other properties of TRSs, e.g., confluence, complexity, and
completion (where it has to be checked for a given TRS $\RR$ and equational
system $\EE$, whether $\RR$ is a convergent TRS that is equivalent to $\EE$). As
a result, there are several tools which support CPF, namely the termination
tools
\aprove~\cite{aprove},
\cime~\cite{a3pat-cime},
\matchbox~\cite{matchbox}, and
\tttt~\cite{ttt2};
the complexity tools
\cat~\cite{cat} and \tct~\cite{tct};
the confluence tool \csi~\cite{csi};
and the completion tools \kbcv~\cite{kbcv} and \mkbtt~\cite{mkbtt}.

CPF is an XML format. As an example proof, in \rFI{xml} we provide the internal representation of
the completion proof of \rSC{example}.
\begin{figure}
\begin{lstlisting}[numbers=left]
<?xml version="1.0"?>
<?xml-stylesheet type="text/xsl" href="cpfHTML.xsl"?>
<certificationProblem xmlns:xsi="http://www.w3.org/2001/XMLSchema-instance"
  xsi:noNamespaceSchemaLocation="cpf.xsd">
  <input>
    <completionInput>
      <equations> ... </equations>
      <trs> ... </trs>
    </completionInput>
  </input>
  <cpfVersion>2.1</cpfVersion>
  <proof>
    <completionProof>
      <wcrProof>
        <joinableCriticalPairsAuto/>
      </wcrProof>
      <trsTerminationProof>
        <ruleRemoval>
          <orderingConstraintProof> ... </orderingConstraintProof>          
          <trs> ... </trs>          
          <trsTerminationProof>
            <rIsEmpty/>
          </trsTerminationProof>
        </ruleRemoval>
      </trsTerminationProof>
      <equivalenceProof>
        <subsumptionProof>
          <ruleSubsumptionProof>
            <rule> ... </rule>
            <conversion> ... </conversion> 
          </ruleSubsumptionProof>
          ...
          <ruleSubsumptionProof> ... </ruleSubsumptionProof>
        </subsumptionProof>
      </equivalenceProof>
    </completionProof>
  </proof>
  <origin>
    <proofOrigin> ... </proofOrigin>
    <inputOrigin> ... </inputOrigin>
  </origin>
</certificationProblem>
\end{lstlisting}
\caption{Internal representation of completion proof of \rSC{example}}\label{xml}
\end{figure}
Each CPF file consists of a single \lstinline!<certificationProblem>! element which 
always has four children: 
the input (lines 5--10), the CPF version number
(line 11), the proof (lines 12--37), and meta information (lines 38--41) which may contain
tool name, configuration of the tool, source of the problem, etc. 

The structure of each proof within CPF is that of an inference tree: each applied technique 
has to contain subproofs for its subgoals and may contain additional information.
For example, the main completion proof in lines 13--36 just contains the three subproofs 
for local confluence (\lstinline!<wcrProof>! stands for weak-Church-Rosser property, another name 
for local confluence), termination, and
equivalence, but no further information; the termination technique of rule removal requires
the additional information of the used well-founded order (line 19), the remaining
TRS (line 20), and the termination proof for the remaining TRS (lines 21--23).

Choosing XML instead of a plain text format was possible as termination
proofs are relatively small. So, the additional size-overhead of XML documents does
not play such a crucial role as it might have played for (large) 
unsatisfiability proofs for SAT or SMT. Similarly, also for other properties like
confluence and complexity, the size overhead never was a problem. Only for completion proofs,
we first encountered problems with too large CPF files of several hundred megabytes. However,
the main problem was not the overhead due to XML, but an exponential blowup when representing
a graph with sharing as fully expanded tree. Once we integrated sharing for these kinds of 
proofs, again the XML files became reasonably sized. As an example for this sharing, 
note that some of the dots ($\ldots$) in \rFI{fig:proof} (or equivalently, some of the
\lstinline!<ruleSubsumptionProof>! elements in lines 28--33 of \rFI{xml}) 
represent intermediate rules which are
not present in the final TRS, but which can be used to derive other conversions.

Using XML has several advantages:
it is easy to generate, since  often programming languages directly offer libraries
for XML processing. Even before certifiers can check the generated proofs, 
standard XML programs can be employed to check whether a CPF file respects the required
XML structure. Finally, it was easy to write a pretty printer to obtain
human readable proofs from CPF files. This pretty printer is written
as an XSL transformation (\texttt{cpfHTML.xsl}), 
so that a browser directly renders CPF proofs in a human readable way:
it just needs to be activated by adding a suitable processing instruction like line 2 of
\rFI{xml}.
Since this pretty printer is freely available,
in principle it is no longer required for tool authors to write their
own human readable proof output: an export to CPF completely suffices. A problem
might occur 
if the tool uses some techniques that are not yet covered by CPF, but then
it is still easily possible to extend the existing pretty printer.

So, after the invention of CPF, the workflow and required proof export
routines for certification have changed from the situation depicted in
\rFI{before} to the less convoluted of \rFI{after}.

\begin{figure}
\centering
\begin{tikzpicture}[node distance = 0.8cm]
\node[tool] (1) {tool 1};
\node[tool, below of = 1] (2) {tool 2};
\node[tool, below of = 2] (3) {tool 3};
\node[below of = 3] (n) {$\vdots$};
\node[cert, right of = 1,xshift=3cm] (a) {human readable};
\node[cert, below of = a] (b) {CPF};
\node[toolc, right of = a, xshift = 3cm] (B) {\colrain};
\node[toolc, below of = B] (C) {\coccime};
\node[toolc, below of = C] (D) {\isaceta};
\foreach \x in {1,2,3}
   {\draw (\x.east) -- (b.west);
    \draw[dashed] (\x.east) -- (a.west);}
\foreach \x in {B,C,D}
   {\draw (b.east) -- (\x.west);}
\draw[->] (b) -- (a);
\end{tikzpicture}
\caption{\label{after}Certification of proofs using CPF}
\end{figure}
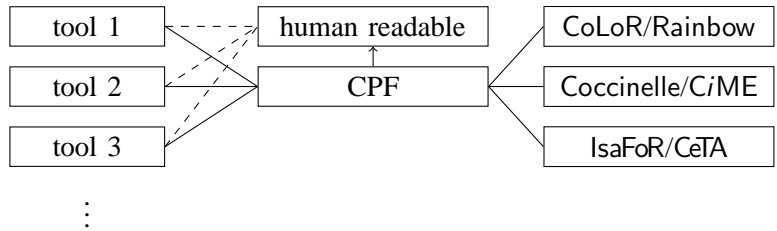

Note that CPF also allows us to represent partial proofs:
The fact that CPF does not support all existing and future
proof methods is reflected by allowing \emph{assumptions}, \emph{unknown proof steps}, and 
\emph{unknown properties}.

Assumptions are useful for being able to specify and certify partial proofs.
For example, even in a failed proof attempt to prove termination of some TRS
$\RR$, several steps may have been applied to simplify $\RR$ into $\RR'$ before
no further progress was possible.  With assumptions it is possible to give a
CPF certificate containing all of the initial steps from $\RR$ to $\RR'$ and
finish the proof by a termination assumption on $\RR'$.  Although such a
certificate does not justify to conclude termination of $\RR$, it can still be
useful to find bugs in the tool: all steps that have been performed are
checkable by a certifier.
Taking this approach to the extreme, we can check individual proof steps of tools by
creating certificates that specify one proof step and finish all subgoals by assumption.
In this way, we added support for \emph{online-certification} to \aprove: when
enabled, then during proof search in \aprove, every single proof step is immediately exported 
into CPF and checked by \ceta, no matter whether the step contributes to the final proof or not.
In this way some bugs in \aprove have been revealed which have not been detected for years.

Unknown proof steps are a generalization of assumptions which may be arbitrary
implications like $P_1 \implies \dots \implies  P_n \implies P_0$ (for
assumptions choose $n = 0$).  However, unknown proof steps serve a different
purpose. Whereas assumptions are there to support certification of partial
proofs (e.g., when an automated tool could not derive the desired property), the
purpose of unknown proof steps is to be able to specify proof steps which have
not been formally defined within CPF (e.g., to specify an application of a
method that has not even been published yet). To this end, each unknown proof
step has to be accompanied by a short textual description.  The main benefit of
unknown proof steps is the ability to certify large parts of \emph{every}
generated proof of some tool, even if it uses some techniques that are unknown
to the certifier.
\begin{example}
Consider the following confluence proof where we assume that
$\succ_1$ is some new reduction order (a well-founded order with further properties) 
which is not yet available in CPF.
\begin{enumerate}
\item
Split the input TRS $\RR$ into the signature disjoint systems $\RR_1$ and $\RR_2$.
Then it suffices to prove confluence of $\RR_1$ and $\RR_2$ separately.

\item
Conclude confluence of $\RR_1$ since it is orthogonal.

\item
Ensure confluence of $\RR_2$ by demanding termination and local confluence.

\item
For termination of $\RR_2$ some rules may be removed due to $\succ_1$, resulting
in $\RR_3$.

\item
Termination of $\RR_3$ is concluded using a reduction order $\succ_2$.
\end{enumerate}
Whereas steps 2, 3, and 5 can be stated in CPF, currently modularity of
confluence (step 1) and step~4 are not supported. However, for both steps the
tool may insert unknown proof steps into the certificate. More precisely,
``modularity of confluence'' (confluence of $\RR_1$ and $\RR_2$ implies
confluence of $\RR$) for step~1 and ``new reduction order $\succ_1$''
(termination of $\RR_3$ implies termination of $\RR_2$) for step~4. Then a
certifier can still check steps 2, 3, and 5 and detect potential problems in
them.
\end{example}

Unknown properties constitute an orthogonal extension that allows us to generate
certificates even for proofs relying on intermediate properties that are not yet
part of CPF. The certifier can just ignore these unknown properties and check
only those parts of the proof that have been specified within CPF.

We conclude this section by giving a complete list of techniques that are
currently supported by CPF. There are certificates for termination and nontermination of
(relative) TRSs and dependency pair problems.  Regarding confluence, CPF
supports certificates for local confluence, confluence, non-confluence, and
completion proofs. Furthermore, CPF supports certificates for the equality and
inequality of two terms with respect to a given set of equations. Another kind
of certificates covers complexity (derivational and runtime) of TRSs. Finally,
there is support for certificates about quasi-reductiveness
of conditional rewriting.

Concerning the individual techniques, 
currently CPF supports several classes of reduction orders
(in alphabetical order): 
argument filters \cite{AG00},
matrix orders \cite{EWZ08},
polynomial orders over several carriers \cite{L79,arctic,rational},
recursive path orders \cite{D87},
the Knuth-Bendix order \cite{KB70}, and 
SCNP reduction orders \cite{CFGSK10}.
Moreover, the techniques of 
dependency graph decomposition \cite{AG00}, 
dependency pairs \cite{AG00,dp-framework}, 
dependency pair transformations (instantiation, narrowing, rewriting) \cite{AG00,dp-framework},
loops,
non-looping nontermination \cite{EEG12},
matchbounds \cite{matchbounds},
root-labeling \cite{root-labeling}, 
rule removal \cite{L79,HM07},
semantic labeling and unlabeling \cite{Z95}, 
size-change termination \cite{LJBA01,AAECC05}, 
string reversal, 
the subterm criterion \cite{HM07}, 
switching to innermost termination \cite{G95},
uncurrying \cite{UncurryHMZ,UncurryST}, 
and usable rules \cite{AG00,U01,dp-framework}
are supported.

Confluence proofs are supported either directly via orthogonal or strongly closed
and linear TRSs, or via Newman's
lemma which requires local confluence and termination. The former by joinability
of critical pairs and the latter reusing all the available machinery for
termination techniques. For non-confluence CPF admits the syntactic criteria mentioned
in \cite{csi}, the tree-automata based techniques of \cite{FT14}, and the methods
based on interpretations and orders of \cite{Aoto13}.

The only technique for quasi-reductiveness of conditional rewrite systems 
is the unraveling transformation in
combination with a termination certificate.

\section{Design Decisions}

In order to gain a wide acceptance by  certifiers as well as automated tools,
representative members of the community have been integrated in the design
process of CPF. 

A main decision was that CPF should provide enough information for all three
certifiers. Currently, there are some elements in CPF that are completely
ignored by some certifier, which in turn are essential for another one. 

After the theoretically required amount of information has been identified,
usually no further details are required in CPF (which eases the generation of
certificates on the side of automated tools).  One exception is that enough
information must be provided to guarantee determinism of each proof step.

\begin{example}
A standard technique to prove termination of a TRS $\RR$ is to 
remove rules by using reduction orders \cite{L79,HM07}.
If the reduction order $\succ$ is provided, 
then usually the result is clear: it is the the remaining TRS $\RR \setminus {\succ}$. 
So in principle, in CPF it should be sufficient to provide $\succ$. However,
since there are several variants of reduction orders and since some reduction
orders---like polynomial orders---are undecidable, it is unclear how $\succ$ is exactly
defined or how it is approximated. To be more concrete, if a polynomial interpretation
over the naturals is provided such that the left-hand side $\ell$ evaluates 
to $p_\ell = x^2 + 1$ and the 
right-hand side $r$ to $p_r = x$, then some approximations can only detect
$\ell \succsim r$ whereas others deliver $\ell \succ r$.
\end{example}

To avoid such problems in CPF, for rule removal and similar ambiguous techniques it is required that the certificate contains enough information to precisely compute each intermediate
proof obligation. As an example, for rule removal it is sufficient to specify either the
removed rules or the remaining rules explicitly, cf.\ line 20 in \rFI{xml}.

An alternative way to achieve determinism is to explicitly demand that the
certificate provides the exact variant or approximation of the reduction order
that is employed, so that the certifier can recompute the same result.
However, this
alternative has the disadvantage that every variant or approximation has to
be exactly specified and even worse, a certifier has to provide algorithms
to compute all variants of reduction orders that are used in termination tools. 
In contrast, with the current solution the certifiers can just implement one
(powerful) variant / approximation of a reduction order. Then during certification
it must just be ensured that all removed rules are strictly decreasing w.r.t.\ $\succ$ (and
the remaining TRS has to be weakly decreasing w.r.t.\ $\succsim$). 

Note that determinism of each proof step is also important for an early detection
of errors. Otherwise, it might happen that a difference in the internal proof 
state of an automated tool and a certifier remains 
undetected for several proof steps. And then errors are reported in proof steps 
which are perfectly okay.

\begin{example}
Let $\RR$ be a TRS consisting of the three rules A, B, and C.
Now consider the following wrong proof by a termination tool:
\begin{enumerate}
\item
Apply rule removal using some reduction order $\succ_1$ to remove rules A and B.
(At this point assume that the termination tool contains a bug, namely that
actually $\succ_1$ would only justify to remove B, but not A.)

\item
Find some other reduction order $\succ_2$ which can remove C.
\item
Conclude termination as there are no rules left.
\end{enumerate}

When we just apply the same techniques during certification without checking
intermediate results, then we first apply rule removal with $\succ_1$ which only
removes B; then we apply rule removal with $\succ_2$, which removes C; and
finally, we report the error that there are rules left, namely A.  This
illustrates that instead of detecting the error in the first step of the tool,
an error pops up in the final step of certification, although the final step in
the termination tool is perfectly fine.
\end{example}

Minor design decisions had to be made for all the supported 
techniques,
e.g., the exact names and the exact representation of relevant parameters, etc.
For these decisions, usually the person who wanted to add a new technique to CPF was
asked to provide a proposal. This proposal was then integrated into a development 
version of CPF and put under discussion on the CPF mailing list.\footnote{%
\href{mailto:cpf@informatik.uibk.ac.at}{\nolinkurl{cpf@informatik.uibk.ac.at}}}
Comments during
the discussion were integrated in the proposal, 
and after the discussion has stopped, the modified proposal was
integrated into the official CPF version.

\section{Conclusion}

We have presented the CPF format, an XML format that allows us to express
several kinds of proofs related to term rewrite systems in a machine readable
format, thereby enabling certification.  So far, CPF is used by several
automated tools and certifiers. New tools are of course welcome to enable CPF
support as well.

Concerning future work, besides the addition of existing techniques that are
currently not supported by any certifier, there is the never ending story of
integrating new proof techniques into CPF.
Moreover, there might be some restructuring of CPF necessary to support very
large proofs. For example, indexing of terms and rules might allow us to
significantly reduce the size of proofs.
Furthermore, for rule removal techniques, CPF might be changed in such a way
that removed rules have to be provided instead of remaining
ones.\footnote{When the remaining rules are stated in the certificate, several
steps of removing a single rule require quadratic size, whereas stating the
removed rules, the size of the overall proof is linear.}
This change would also allow us to represent rule removal techniques for
termination and relative termination in the same way, which in turn would allow
us to merge proof techniques for termination and relative termination. 

However, some of above changes are non-conservative and thus require adaptations
of the proof generating tools and certifiers. Therefore, we believe that it
should be discussed thoroughly by the community whether such changes should be
made.\footnote{In principle it is also possible to integrate changes to some
existing proof technique as a new technique which co-exists with its original
obsolete version. In this way, changes can be made conservatively. But then any
clean-up of outdated techniques will be a non-conservative change.} Some
non-conservative changes have already been made when switching from CPF version
1.0 to version 2.0. 

Another interesting point is tool collaboration, i.e., either on the producing
side (e.g., several tools producing a single certificate) or on the consuming
side (e.g., several certifiers taking care of different parts of a certificate).
While the former is already done in practice, e.g., confluence tools rely on
termination tools and reuse thusly obtained certificates when producing their
own CPF output, the latter poses some trustability issues. Even if subproofs are
rigorously certified by different certifiers there has to be a trustworthy way
to combine those results since there is no guarantee that different certifiers
are based on the same underlying semantics. We leave this crucial issue as
future work.

\bibliographystyle{eptcs}
\bibliography{references}

\end{document}